\pdfoutput=1
\documentclass[11pt, a4paper]{article}
\usepackage{jcappub.local}

\usepackage{subfigure}
\usepackage[subfigure]{graphfig}
\usepackage{color}
\usepackage[toc,page]{appendix}

\usepackage[section]{placeins}
\makeatletter
\AtBeginDocument{%
  \expandafter\renewcommand\expandafter\subsection\expandafter{%
    \expandafter\@fb@secFB\subsection
  }%
}
\makeatother

\title{Inflation with Massive Vector Fields}

\author[a]{Junyu Liu}
\author[b]{Yi Wang}
\author[b]{Siyi Zhou}
\affiliation[a]{University of Science and Technology of China, \\
Hefei, Anhui 230026, P.R.China}
\affiliation[b]{Department of Physics, The Hong Kong University of Science and Technology,\\
Clear Water Bay, Kowloon, Hong Kong, P.R.China}
\emailAdd{junyu@mail.ustc.edu.cn}
\emailAdd{phyw@ust.hk}
\emailAdd{szhouah@ust.hk}

\abstract{We investigate the coupling between the inflaton and massive vector fields. All renormalizable couplings with shift symmetry of the inflaton are considered. The massive vector can be decomposed into a scalar mode and a divergence-free vector mode. We show that the former naturally interacts with the inflaton and the latter decouples at tree level. The model in general predicts $f_{NL}^\mathrm{equil} = \mathcal{O}(1)$, while in some regions of the parameter space large non-Gaussianity can arise.}

\begin{document}

\maketitle

\section{Introduction} \label{sec:intro}
Inflation is the most promising paradigm for the study of the early universe~\cite{Guth:1980zm,Linde:1981mu,Albrecht:1982wi}. Despite its history and success, the detailed dynamics of inflation is still unknown. Especially, we are not sure if inflation is driven by one field, multiple fields, or quasi-single field. In the case of quasi-single field inflation, the isocurvatons have mass of order $H$~\cite{Chen:2009we,Chen:2009zp}. The model is later extended to higher mass scales~\cite{Chen:2012ge, Pi:2012gf}. Similar observational signatures also arise in CFT with anomalous operator dimensions~\cite{Green:2013rd}.

In previous studies, the model of quasi-single field inflation is investigated in the case where the isocurvatons are scalar fields. It is interesting to extend the study to higher spin fields. On the one hand, on colliders, we have found a full spectrum of spin-$1/2$ and spin-$1$ fields but only one spin-$0$ field. Thus it is natural to expect the existence of those higher spin fields at relevant scales during inflation~\cite{Ford:1989me,Maleknejad:2011jw,Solomon:2013iza}. On the other hand, one important motivation of quasi-single field inflation, supersymmetry~\cite{Baumann:2011nk}, relates particles with different spins. Thus a systematic study of the field spectrum helps for understanding the UV completion of quasi-single field inflation.

In the present work, we study the coupling between the inflaton and a massive vector field. Although we are motivated to study the case where the vector field mass is $M\sim H$, the calculation also holds for a general mass. The mass of the vector field would either be fundamental, or from symmetry breaking at higher energy scales compared to inflation.

This paper is organized as follows:
In Section 2, we study the possible interaction terms in this model. We list all possible interaction terms and identify a subset of terms which are relevant for our study.
In Section 3, we introduce a decomposition method and show that in this case, the coupling to the vector field can be reduced to the coupling to a scalar.
In Section 4, we calculate the power spectrum of this model by rotating the field basis. At leading non-trivial order, the result is reproduced by an in-in formalism calculation, which we present in the appendix.
In Section 5,  we study the non-Gaussianity of this model.
In Section 6, we consider the multi-vector field case.
In Section 7, we conclude and discuss some possible future directions.

\section{The renormalizable terms}
In this section, we introduce our setup for the model involving a scalar field as the inflaton and a vector field as an isocurvaton.
We first summarize all the possible terms in the Lagrangian in Table \ref{term}. Here we only consider the terms with a shift symmetry in $\varphi$, and the resulting equation of motion is at most second order.
\begin{table}[htbp]
\begin{center}
    \begin{tabular}{| c | c |}
    \hline
   dimension & Possible terms in the Lagrangian \\ \hline
   d=2 &  $ A^2$ \\ \hline
   d=3 & $ A^\mu\partial _ \mu\varphi$ \\ \hline
   d=4 & $A^4, F_{\mu \nu}F^{\mu \nu}, \varphi A^\mu\partial _ \mu \varphi, \partial_\mu\varphi \partial^\mu\varphi$ \\   \hline
    \end{tabular}
\caption{This table summarizes all the possible terms in the lagrangian when we consider inflation with a scalar and a massive vector field. Here we require the scalar field to have a shift symmetry. Also, terms which lead to higher than second order equation of motion is not considered.}\label{term}
\end{center}
\end{table}

To summarize our setup, we can write down the Lagrangian of our model as
\begin{equation}
\mathcal{L}=\sqrt{-g} \left[  -\frac{1}{2}\partial^\mu\varphi\partial_\mu\varphi-\frac{1}{4}F^{\mu\nu}F_{\mu\nu}-\frac{1}{2}M^2 A^{\mu}A_{\mu}+m A^{\mu}\partial_{\mu}\varphi+\beta\varphi A^\mu\partial_\mu\varphi+\kappa (A^{\mu}A_{\mu})^2-V_{sr}(\varphi)    \right]~,
\end{equation}
where $V_{sr}(\varphi)$ is the usual slow-roll potential. The signature of the metric is $(-1,1,1,1)$. $M$ is the mass of the vector field and $m$, $\beta$ and $\kappa$ are constants.

\section{Decomposition}
In this section, we decompose the massive vector field into a transverse part and a scalar part. We consider the transformation
\begin{equation}
A_{\mu}=\tilde{A}_\mu+\partial_\mu \tilde{\sigma}~,
\end{equation}
such that
\begin{equation}
D^\mu \tilde{A}_\mu = 0~.
\end{equation}
Here the $\tilde{\sigma}$ field does not have mass dimension.
We can perform a field redefinition $\sigma = M\tilde{\sigma}$.

Then the lagrangian takes the form
\begin{align}
\mathcal{L} = \sqrt{-g}\Big[ & { -\frac{1}{2}\partial^\mu\varphi\partial_\mu\varphi-\frac{1}{4}\tilde{F}^{\mu\nu}\tilde{F}_{\mu\nu}-\frac{1}{2}\partial^\mu\sigma\partial_\mu\sigma+\alpha \partial^\mu\sigma\partial_{\mu}\varphi}  \nonumber\\
 & { + \omega \varphi \partial^\mu\sigma\partial_\mu\varphi+\gamma(\partial^\mu\sigma\partial_\mu\sigma)^2 +\textrm{(terms with $\tilde{A}$ and $\tilde{\sigma}$)}  -V_{sr}(\varphi)   } \Big] ~,\\
\end{align}
where $\alpha=\frac{m}{M}$, $\gamma = \frac{\kappa}{M^4}$ and $\omega = \frac{\beta}{M}$. Here the $-\frac{1}{4}F^{\mu\nu}F_{\mu\nu}$ term is invariant under this transformation and after the transformation it does not couple to the $\varphi$ field, so we do not consider it (for the same reason, $\epsilon^{\mu\nu\rho\lambda}F_{\mu\nu}F_{\rho\lambda}$ is not relevant for our purpose). We do not consider either the coupling between $\tilde{A}^\mu$ and $\partial_\mu\tilde{\sigma}$, which only contributes to higher order corrections in the perturbation theory.

\section{Power spectrum and non-Gaussianities}
We start from the quadratic part of the Lagrangian
\begin{equation}
\mathcal{L}=\sqrt{-g} \left[ -\frac{1}{2}{{\partial }^{\mu }}\varphi {{\partial }_{\mu }}\varphi -\frac{1}{2}{{\partial }^{\mu }}\sigma {{\partial }_{\mu }}\sigma +\alpha{{\partial }^{\mu }}\varphi {{\partial }_{\mu }}\sigma - V_{sr}(\varphi)  \right]~.
\end{equation}
It is convenient to rotate the fields to remove the mixing terms. Before doing that, it is useful to note that the value of $\alpha$ is constrained. We can write down the kinetic matrix of Lagrangian as
\begin{equation}
\mathcal{L}=-\frac{1}{2}\sqrt{-g}\left( \begin{matrix}
   {{\partial }^{\mu }}\varphi  & {{\partial }^{\mu }}\sigma   \\
\end{matrix} \right)\left( \begin{matrix}
   1 & -\alpha   \\
   -\alpha  & 1  \\
\end{matrix} \right)\left( \begin{matrix}
   {{\partial }_{\mu }}\varphi   \\
   {{\partial }_{\mu }}\sigma   \\
\end{matrix} \right)~.
\end{equation}
In order to avoid the ghost, we should make sure that the kinetic matrix is positive-definite, namely
\begin{equation}
\alpha^2<1~.
\end{equation}
To solve the model, one can rotate field basis as
\begin{align}
&\varphi =a \psi +b \chi  ~,\\
&\sigma =c \psi +d \chi ~,\nonumber
\end{align}
where $\psi$ and $\chi$ are the redefined fields, and the coefficients $a,b,c,d$ are constants.

It is convenient to require the kinetic term of the new fields to be the same as free-field case, with the mixing term being eliminated. After solving the equations for $a,b,c$ and $d$, we have the relation:
\begin{equation}
{{a }^{2}}+{{b }^{2}}=\frac{1}{1-{{\alpha}^{2}}}~.
\end{equation}
Now we compute the two point correlation function of the fluctuation of the $\varphi$ field
\begin{equation}
\left\langle {{\delta\varphi }^{2}} \right\rangle ={{a }^{2}}\left\langle {{\delta\psi }^{2}} \right\rangle +{{b }^{2}}\left\langle {{\delta\chi }^{2}} \right\rangle +ab \left\langle \delta\psi \delta\chi  \right\rangle +ab \left\langle \delta\chi \delta\psi  \right\rangle~,
\end{equation}
where ${\delta{\varphi }^{2}}$ is understood as ${\delta{\varphi }_{{{\mathbf{k}}_{\mathbf{1}}}}}(t){{\delta\varphi }_{{{\mathbf{k}}_{\mathbf{2}}}}}(t)$ here, which is the same for other fields. Note that there is no interaction between $\delta\psi$ and $\delta\chi$, we have
\begin{equation}
\left\langle {{\delta\psi }^{2}} \right\rangle =\left\langle {{\delta\chi }^{2}} \right\rangle =\frac{{{H}^{2}}}{2k_{1}^{3}}{{(2\pi )}^{3}}{{\delta }^{(3)}}({{\mathbf{k}}_{\mathbf{1}}}+{{\mathbf{k}}_{\mathbf{2}}})~,
\end{equation}
\begin{equation}
\left\langle \delta\psi \delta\chi  \right\rangle =\left\langle \delta\chi \delta\psi  \right\rangle =0~.
\end{equation}
Finally
\begin{equation}
\left\langle {{\delta\varphi }^{2}} \right\rangle =\frac{1}{1-{{\alpha}^{2}}}\frac{{{H}^{2}}}{2k_{1}^{3}}{{(2\pi )}^{3}}{{\delta }^{(3)}}({{\mathbf{k}}_{\mathbf{1}}}+{{\mathbf{k}}_{\mathbf{2}}})~.
\end{equation}

When $\alpha$ is small, we can expand around $\alpha=0$ to get
\begin{equation}
\left\langle {{\delta\varphi }^{2}} \right\rangle =\frac{{{H}^{2}}}{2k_{1}^{3}}{{(2\pi )}^{3}}{{\delta }^{(3)}}({{\mathbf{k}}_{\mathbf{1}}}+{{\mathbf{k}}_{\mathbf{2}}})(1+{{\alpha}^{2}})~.
\end{equation}
This result agrees with a direct in-in calculation which treats the mixing of $\varphi$ and $\sigma$ as interactions. We present the in-in calculation in the appendix.

To relate the inflaton perturbation to the curvature perturbation $\zeta$, we have to identify the reheating surface. If the reheating surface depends both on the inflaton and the vector directions, the isocurvature fluctuation would convert into the curvature fluctuation via multi-brid mechanism \cite{Sasaki:2008uc, Huang:2009vk}. However, we note that if we choose the rotation of this form
\begin{align}\label{naturalrotation}
&\varphi =\frac{1}{\sqrt{1-{{\alpha }^{2}}}}\psi ~,\\
&\sigma =\frac{\alpha }{\sqrt{1-{{\alpha }^{2}}}}\psi +\chi ~,\nonumber
\end{align}
a natural ending mechanism of our inflationary scenario only involves the $\psi$ field (which is proportional to the $\varphi$ field). This is because $V_{sr}$ depends only on $\psi$ but not $\chi$. Thus no matter the end of inflation is triggered by a given field value, or a given Hubble time, the value of $\psi$ domains the end of inflation.

Making use of the time delay formula $\zeta = - H \delta\psi / \dot\psi$, we get
\begin{align}
  P^\zeta_k = \frac{H^2}{8\pi^2\epsilon M_p^2}~.
\end{align}
In other words, the power spectrum is formally unchanged, if $\epsilon$ is defined via the Hubble parameter
\begin{align}
  \epsilon = - \frac{\dot H}{H^2} \simeq \frac{M_p^2}{2} \left( \frac{1}{V} \frac{dV}{d\psi} \right)^2~,
\end{align}
where at the second half of the equation we have used the slow roll equations of motion of the $\psi$ field. The same fact can be found by noting that, although $\delta\varphi$ is rescaled by a factor of $1/\sqrt{1-\alpha^2}$, $\dot\varphi$ is also rescaled by the same factor. So we get the same result when using $\zeta = - H \delta\varphi / \dot\varphi$ to calculate $\zeta$. However, in terms of the original model, we have
\begin{align}
  \epsilon_V = \frac{M_p^2}{2} \left( \frac{1}{V} \frac{dV}{d\varphi} \right)^2
  = \epsilon \left( \frac{d\psi}{d\varphi} \right)^2 = (1-\alpha^2)\epsilon~.
\end{align}
As a result, with the convenient $\epsilon_V$ parameter, the power spectrum is expressed as
\begin{align}
  P^\zeta_k = \frac{(1-\alpha^2) H^2}{8\pi^2\epsilon_V M_p^2}~.
\end{align}

Now we move on to include higher order terms in the Lagrangian:
\begin{equation}
\mathcal{L}=\sqrt{-g} \left[ -\frac{1}{2}{{\partial }^{\mu }}\varphi {{\partial }_{\mu }}\varphi -\frac{1}{2}{{\partial }^{\mu }}\sigma {{\partial }_{\mu }}\sigma +\alpha {{\partial }^{\mu }}\varphi {{\partial }_{\mu }}\sigma +\gamma {{\left( {{\partial }^{\mu }}\sigma {{\partial }_{\mu }}\sigma  \right)}^{2}} - V_{sr}(\varphi)  \right]~,
\end{equation}
where $\alpha =\frac{m}{M}$, $\gamma =\frac{\kappa}{M^4}$ are coupling constants (The addition Lagrangian is $\mathcal{L}_4=\kappa A^4$). The $\varphi \partial^\mu\sigma\partial_\mu\varphi$ term gives only slow roll suppressed contributions to non-Gaussianity thus we are not considering it here. The reason is explained at the end of this section.

We first consider the background behavior of this Lagrangian. The classical Euler-Lagrangian equations give us the classical equations of motion
\begin{equation}
3H\dot{\varphi}+\ddot{\varphi}+V_{sr}(\varphi)=\alpha (3H\dot{\sigma}+\ddot{\sigma})~,
\end{equation}
\begin{equation}
3H\dot{\sigma}+\ddot{\sigma}-3H\alpha\dot{\varphi}-\alpha\ddot{\varphi}+12\gamma\dot{\sigma}^2(H\dot{\sigma}+\ddot{\sigma})=0~.
\end{equation}
With the field rotation (\ref{naturalrotation}), ignoring the cross term from the bi-quadratic coupling of vector and the kinetic term of $\chi$ field, we have the new Lagrangian of new fields
\begin{equation}
\mathcal{L}=-\frac{1}{2}\sqrt{-g}{{\partial }^{\mu }}\psi {{\partial }_{\mu }}\psi -\sqrt{-g}{{V}_{sr}}(\theta \psi )+\sqrt{-g}\Theta {{({{\partial }^{\mu }}\psi {{\partial }_{\mu }}\psi )}^{2}}~,
\end{equation}
where
\begin{equation}
\theta =\frac{1}{\sqrt{1-{{\alpha }^{2}}}}~,
\end{equation}
\begin{equation}
\Theta =\gamma \alpha^4 {{\theta }^{4}}~.
\end{equation}
Now we need to consider correlators of $\psi$. The calculation of this model follows from \cite{Chen:2006nt}. We denote
\begin{equation}
{{\mathcal{L}}}=\sqrt{-g}P(X,\psi )~,
\end{equation}
where
\begin{equation}
P(X,\psi )=X-{{V}_{sr}}(\theta \psi )+4\Theta X^2=X+DX^2-{{V}_{sr}}(\theta \psi )~,
\end{equation}
\begin{equation}
X=-\frac{1}{2}{{\partial }^{\mu }}\psi {{\partial }_{\mu }}\psi~,
\end{equation}
\begin{equation}
D=4\Theta=\frac{4\kappa }{{{M}^{4}}} \left( \frac{m}{M}  \right)^4  {{\left( 1-{{\left( \frac{m}{M} \right)}^{2}} \right)}^{-2}}~.
\end{equation}
The energy density is
\begin{equation}
\rho =3M_{p}^{2}{{H}^{2}}=2X{{P}_{,X}}-P=X+3D{{X}^{2}}+{{V}_{sr}}(\theta \psi )~.
\end{equation}
Define the sound speed as
\begin{equation}
c_{s}^{2}=\frac{{{P}_{,X}}}{{{P}_{,X}}+2X{{P}_{,XX}}}~.
\end{equation}
We define another useful parameter
\begin{equation}
u=\frac{1}{c_s^2}-1~.
\end{equation}
The leading order power spectrum of this model is given by
\begin{equation}\label{5.16}
P_{k}^{\zeta }=\frac{{{H}^{2}}}{8{{\pi }^{2}}M_{p}^{2}{{c}_{s}}\epsilon }~.
\end{equation}

It is very useful to define the following two parameters when we study the non-Gaussianity of this model
\begin{equation}
\Sigma =X{{P}_{,X}}+2{{X}^{2}}{{P}_{X,X}}~,
\end{equation}
\begin{equation}
\lambda ={{X}^{2}}{{P}_{X,X}}+\frac{2}{3}{{X}^{3}}{{P}_{X,X,X}}~.
\end{equation}
Define some parameters that are more general than slow-roll as
\begin{equation}
\epsilon =-\frac{{\dot{H}}}{{{H}^{2}}}=\frac{X{{P}_{,X}}}{{{H}^{2}}}=\mathcal{O}(\epsilon )~,
\end{equation}
\begin{equation}
\eta =\frac{{\dot{\epsilon }}}{H\epsilon }=\mathcal{O}(\epsilon )~,
\end{equation}
\begin{equation}
s=\frac{{{{\dot{c}}}_{s}}}{H{{c}_{s}}}=\mathcal{O}({{\epsilon }^{2}})~,
\end{equation}
\begin{equation}
l=\frac{{\dot{\lambda }}}{\lambda H}=\mathcal{O}(\epsilon )~.
\end{equation}

The bispectrum is already calculated in~\cite{Chen:2006nt}:
\begin{equation}
\left\langle {{\zeta }_{{{\mathbf{k}}_{\mathbf{1}}}}}{{\zeta }_{{{\mathbf{k}}_{\mathbf{2}}}}}{{\zeta }_{{{\mathbf{k}}_{\mathbf{3}}}}} \right\rangle ={{(2\pi )}^{7}}{{\delta }^{(3)}}({{\mathbf{k}}_{\mathbf{1}}}+{{\mathbf{k}}_{2}}+{{\mathbf{k}}_{3}}){{(P_{K}^{\zeta })}^{2}}\frac{1}{k_{1}^{3}k_{2}^{3}k_{3}^{3}}({{A}_{\lambda }}+{{A}_{c}}+{{A}_{o}}+{{A}_{\epsilon }}+{{A}_{\eta }}+{{A}_{s}})~,
\end{equation}
where
\begin{equation}
K={{k}_{1}}+{{k}_{2}}+{{k}_{3}}~,
\end{equation}
\begin{equation}
{{A}_{\lambda }}=\frac{3}{2}(u-2\frac{\lambda }{\Sigma })\frac{k_{1}^{2}k_{2}^{2}k_{3}^{2}}{{{K}^{3}}}~,
\end{equation}
\begin{equation}
{{A}_{c}}=u(-\frac{1}{K}\sum\limits_{i>j}{k_{i}^{2}k_{j}^{2}}+\frac{1}{2{{K}^{2}}}\sum\limits_{i\ne j}{k_{i}^{2}k_{j}^{3}}+\frac{1}{8}\sum\limits_{i}{k_{i}^{3}})~,
\end{equation}
\begin{equation}
{{A}_{o}}={{A}_{\epsilon }}={{A}_{\eta }}=\mathcal{O}(\epsilon )~,
\end{equation}
\begin{equation}
{{A}_{s}}=\mathcal{O}({{\epsilon }^{2}})~.
\end{equation}
The variables like $H$, $c_s$ and $\epsilon$ are evaluated at the time when the wave number $K$ exits the horizon, namely, all the thing are measured at $\tau_k=\frac{1}{Kc_{sK}}+\mathcal{O}(\epsilon)=\frac{1}{a_k H_K}+\mathcal{O}(\epsilon)$. So the observable contributions are
\begin{equation}
A=\frac{24{{\left( DX \right)}^{2}}}{1+8DX+12{{\left( DX \right)}^{2}}}\frac{k_{1}^{2}k_{2}^{2}k_{3}^{2}}{{{K}^{3}}}+\frac{4DX}{1+2DX}(-\frac{1}{K}\sum\limits_{i>j}{k_{i}^{2}k_{j}^{2}}+\frac{1}{2{{K}^{2}}}\sum\limits_{i\ne j}{k_{i}^{2}k_{j}^{3}}+\frac{1}{8}\sum\limits_{i}{k_{i}^{3}})~.
\end{equation}

When we use the estimator $f_{{NL}}$, conventionally\footnote{Here the sign convention differs from \cite{Chen:2006nt}. We use the convention by \textit{WMAP} and \textit{Planck} for $f_{NL}$.}
\begin{equation}
\left\langle {{\zeta }_{{{\mathbf{k}}_{\mathbf{1}}}}}{{\zeta }_{{{\mathbf{k}}_{\mathbf{2}}}}}{{\zeta }_{{{\mathbf{k}}_{\mathbf{3}}}}} \right\rangle ={{(2\pi )}^{7}}{{\delta }^{(3)}}({{\mathbf{k}}_{\mathbf{1}}}+{{\mathbf{k}}_{2}}+{{\mathbf{k}}_{3}}){{(P_{k}^{\zeta })}^{2}}(\frac{3}{10}){{f}_{\mathrm{NL}}}\frac{k_{1}^{3}+k_{2}^{3}+k_{3}^{3}}{k_{1}^{3}k_{2}^{3}k_{3}^{3}}~,
\end{equation}
at $k_1=k_2=k_3$. Now the measurable contribution is
\begin{equation}
f_{{NL}}^{\lambda }=\frac{5}{81}(u-2\frac{\lambda }{\Sigma })=\frac{80}{81}\frac{{{(DX)}^{2}}}{1+8(DX)+12{{(DX)}^{2}}}~,
\end{equation}
\begin{equation}
f_{{NL}}^{c}=-\frac{35}{108}u=-\frac{35}{27}\frac{DX}{1+2DX}~,
\end{equation}
So
\begin{equation}
{{f}_{{NL}}}=f_{\mathrm{NL}}^{c}+f_{NL}^{\lambda }=-\frac{5}{81}\frac{DX(21+110DX)}{(1+2DX)(1+6DX)}~.
\end{equation}
First, we assume that $D>0$, namely $\kappa>0$. Now it is very easy to show that this function is decreasing when $DX$ increases from zero to positive infinity. It can be seen from the following Figure~\ref{fig:DX1}. The function has a limitation value $-\frac{275}{486}=-0.57$.
\begin{figure}[htbp]
  \centering
  \includegraphics[width=0.7\textwidth]{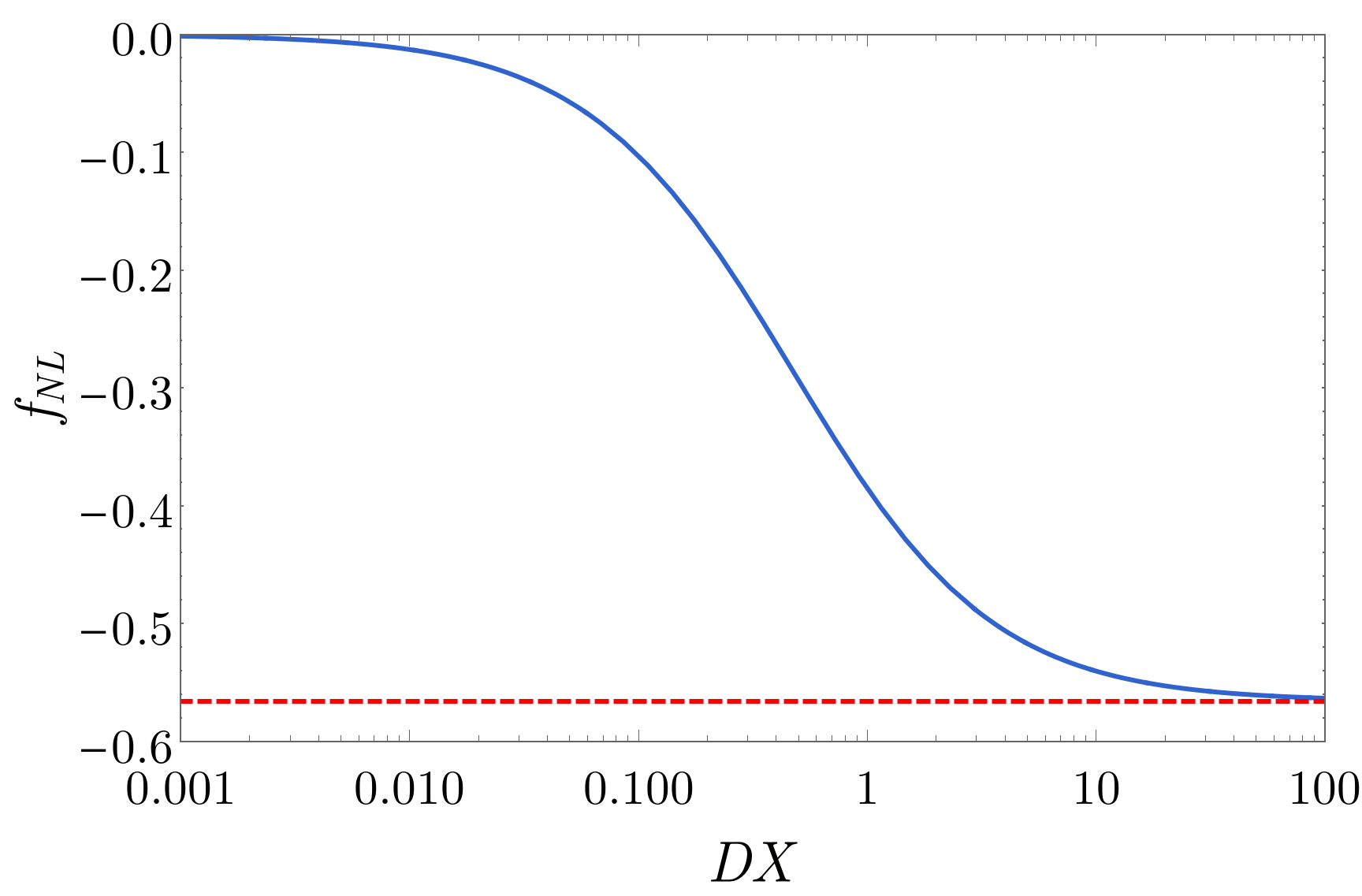}
  \caption{\label{fig:DX1} The function $f_{NL}$ with $D>0$. The blue curve is the graph of function $f_{NL}$, while the red dashed line is its minimal $-\frac{275}{486}$.}
\end{figure}
So we can get the range
\begin{equation}
\left|{{f}_{NL}}\right|\le \frac{275}{486}=0.57~.
\end{equation}

Now we consider the limiting case. When inflation happens, we have $X \sim \epsilon {{H }^{2}}M_{p}^{2}$, so
\begin{equation}
DX\sim D\epsilon {{H }^{2}}M_{p}^{2}\left(\frac{m}{M}\right)^4=\frac{4\kappa }{{{M}^{4}}}\left(\frac{m}{M}\right)^4{{\left( 1-{{\left( \frac{m}{M} \right)}^{2}} \right)}^{-2}}\epsilon {{H}^{2}}M_{p}^{2}~.
\end{equation}
If $\alpha=\frac{m}{M}$ is not too small compared to 1, We can choose $M\sim H$ so $DX$ can be extremely large. In this limit
\begin{equation}
c_{s}^{2}=\frac{1}{3}~,
\end{equation}
\begin{equation}
u=2~,
\end{equation}
\begin{equation}
\frac{\lambda}{\Sigma}=\frac{1}{3}~.
\end{equation}
So the power spectrum is
\begin{equation}
P_{k}^{\zeta }=\frac{\sqrt{3}{{H}^{2}}}{8{{\pi }^{2}}M_{p}^{2}\epsilon }~.
\end{equation}
The three point function amplitude is
\begin{equation}
A=2\frac{k_{1}^{2}k_{2}^{2}k_{3}^{2}}{{{K}^{3}}}+2(-\frac{1}{K}\sum\limits_{i>j}{k_{i}^{2}k_{j}^{2}}+\frac{1}{2{{K}^{2}}}\sum\limits_{i\ne j}{k_{i}^{2}k_{j}^{3}}+\frac{1}{8}\sum\limits_{i}{k_{i}^{3}})~,
\end{equation}
and $\left|f_{NL} \right|$ achieves its maximal
\begin{equation}
\left|{{f}_{NL}}\right|=\frac{275}{486}=0.57~.
\end{equation}

\begin{figure}[htbp]
  \centering
  \includegraphics[width=0.7\textwidth]{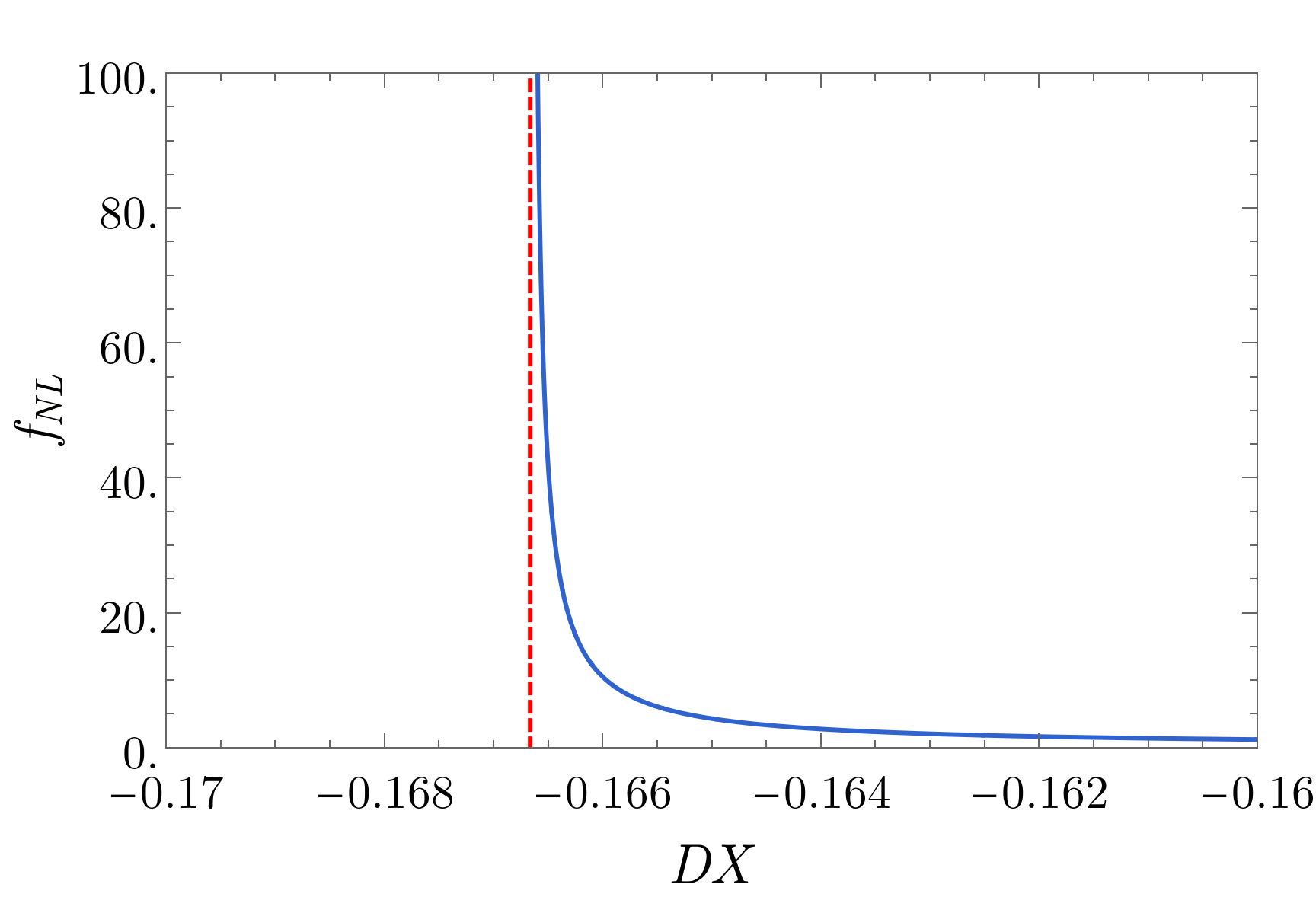}
  \caption{\label{fig:DX2} The $f_{NL}$ is blowing up where $D<0$. The blue curve is the graph of function $f_{NL}$, while the red dashed line is its asymptote $DX=-\frac{1}{6}$.}
\end{figure}

It is also interesting to consider the case where $D<0$, which implies $\kappa<0$. In this case, we can also plot $f_{NL}$ as a function of $DX$. (See Figure~\ref{fig:DX2}.) The ghost free condition requires
\begin{equation}
\epsilon =-\frac{{\dot{H}}}{{{H}^{2}}}=\frac{X{{P}_{,X}}}{{{H}^{2}}}=\frac{3X+6D{{X}^{2}}}{X+3D{{X}^{2}}+{{V}_{sr}}(\theta \psi )}>0~.
\end{equation}
Consider that the slow roll potential ${V}_{sr}$ dominates the energy density of inflation, we have
\begin{equation}
DX>-\frac{1}{2}~.
\end{equation}
Similarly, the absence of tachyonic instability requires
\begin{equation}
c_{s}^{2}=\frac{{{P}_{,X}}}{{{P}_{,X}}+2X{{P}_{,XX}}}=\frac{1+2DX}{1+6DX}>0~.
\end{equation}
Thus
\begin{equation}
DX>-\frac{1}{6}~.
\end{equation}
In Figure.~\ref{fig:DX2}, we show that this model can provide a large amplitude of $f_{NL}$ when $DX\sim-\frac{1}{6}$.

Finally, we comment on the term $\sqrt{-g} w \varphi \partial^\mu \sigma \partial_\mu \varphi$. First, we expect $|w \varphi| < \mathcal{O}(1)$ because otherwise ghost arises in the perturbations, similar to the case of $|\alpha| > 1$. In the parameter region of $|w \varphi| < \mathcal{O}(1)$, this term does not significantly modify the background slow roll dynamics. Still, naive dimensional analysis suggest that this term would produce large non-Gaussianity for natural $w$. However, following the formalism of \cite{Chen:2006nt}, we find that this term does not modify the sound speed, nor the $\lambda / \Sigma$ term. Thus no large non-Gaussianity is generated. This fact can alternatively be observed as follows: The perturbation of this term can be transformed to a term which is proportional to the field equation of motion
\begin{align}
  \sqrt{-g} w \varphi \partial^\mu \sigma \partial_\mu \varphi
  \sim \frac{\alpha}{(1-\alpha^2)^{3/2}}\sqrt{-g} w \psi \partial^\mu \psi \partial_\mu \psi
  \sim - \frac{\alpha}{2(1-\alpha^2)^{3/2}}\sqrt{-g} w \psi^2 \Box \psi~,
\end{align}
where we first neglected irrelevant $\chi$ terms, and then performed integration by parts. Thus one can redefine $\psi$ to cancel this term. The field redefinition only leads to a slow roll suppressed contribution to the local non-Gaussianity thus we are not considering this term here.

\section{Inflation with a tower of massive vectors}

A fundamental vector field may follow from compactification of extra dimensions, or massive string oscillation states. In either case, one gets not only one, but instead a tower of massive states. Following the above motivation, in this section we consider a tower of massive vectors, which is a direct generalization of the previous sections.

We consider the Lagrangian
\begin{equation}
\mathcal{L}=\sum\limits_{i=1}^{N}{\mathcal{L}_{A}^{(i)}}+\sum\limits_{i=1}^{N}{\mathcal{L}_{I}^{(i)}}+{{\mathcal{L}}_{\varphi }}~,
\end{equation}
where
\begin{equation}
\mathcal{L}_{A}^{(i)}=\sqrt{-g}\left( -\frac{1}{4}F_{\mu \nu }^{(i)}{{F}^{\mu \nu (i)}}-\frac{1}{2}M_{i}^{2}A_{\mu }^{(i)}{{A}^{\mu (i)}} \right)~,
\end{equation}
\begin{equation}
\mathcal{L}_{I}^{(i)}=\sqrt{-g}m_i{{\partial }_{\mu }}\varphi {{A}^{\mu (i)}}~,
\end{equation}
\begin{equation}
{{\mathcal{L}}_{\varphi }}=\sqrt{-g}\left( -\frac{1}{2}{{\partial }^{\mu }}\varphi {{\partial }_{\mu }}\varphi -{{V}_{sr}}(\varphi ) \right)~.
\end{equation}
This is the mass tower vector set ${{\left\{ {{A}^{\mu (i)}} \right\}}_{i=1,2...N}}$ with corresponding field strength\\ ${{\left\{ F_{\mu \nu }^{(i)}={{\partial }_{\mu }}A_{\nu }^{(i)}-{{\partial }_{\nu }}A_{\mu }^{(i)} \right\}}_{i=1,2...N}}$, mass ${{\left\{ {{M}_{i}} \right\}}_{i=1,2...N}}$ and coupling constants ${{\left\{ {{m}_{i}} \right\}}_{i=1,2...N}}$.

Decompose the vectors as
\begin{equation}
A_{\mu }^{(i)}=\tilde{A}_{\mu }^{(i)}+{{\partial }_{\mu }}{{\tilde{\sigma }}^{(i)}}~,
\end{equation}
satisfying
\begin{equation}
{{D }^{\mu }}\tilde{A}_{\mu }^{(i)}=0~.
\end{equation}
Then we redefine the $\sigma$ field ${{\sigma }^{(i)}}={{M}_{i}}{{\tilde{\sigma }}^{(i)}}$. We have
\begin{equation}
\mathcal{L}_{I}^{(i)}=\sqrt{-g}\frac{{{m}_{i}}}{{{M}_{i}}}{{\partial }_{\mu }}\varphi {{\partial }^{\mu }}{{\sigma }^{(i)}}~,
\end{equation}
and
\begin{equation}
\mathcal{L}_{A}^{(i)}=\sqrt{-g}\left( -\frac{1}{4}\tilde{F}_{\mu \nu }^{(i)}{{{\tilde{F}}}^{\mu \nu (i)}}-\frac{1}{2}M_{i}^{2}\tilde{A}_{\mu }^{(i)}{{{\tilde{A}}}^{\mu (i)}}-\frac{1}{2}{{\partial }_{\mu }}{{\sigma }^{(i)}}{{\partial }^{\mu }}{{\sigma }^{(i)}} \right)=\mathcal{L}_{{\tilde{A}}}^{(i)}+\mathcal{L}_{\sigma }^{(i)}~.
\end{equation}

Notice that the $\tilde{A}^{(i)}$ modes only couples to themselves. So we can ignore them and effectively write our total Lagrangian as
\begin{equation}
\mathcal{L}=-\sum\limits_{i=1}^{N}{\frac{\sqrt{-g}}{2}{{\partial }_{\mu }}{{\sigma }^{(i)}}{{\partial }^{\mu }}{{\sigma }^{(i)}}}+\sum\limits_{i=1}^{N}{\sqrt{-g}{{\alpha }_{i}}{{\partial }_{\mu }}\varphi {{\partial }^{\mu }}{{\sigma }^{(i)}}}-\frac{\sqrt{-g}}{2}{{\partial }^{\mu }}\varphi {{\partial }_{\mu }}\varphi -\sqrt{-g}{{V}_{sr}}(\varphi )~,
\end{equation}
where ${{\alpha }_{i}}=\frac{{{m}_{i}}}{{{M}_{i}}}$. This is similar to the single vector case. Notice that we still have the constraint,
\begin{equation}
\sum\limits_{i=1}^N\alpha_i^2<1~,
\end{equation}
in order to avoid ghosts.

Now rotate our $N+1$ scalars as
\begin{equation}
\varphi =\frac{1}{\sqrt{1-\sum\limits_{i=1}^{N}{\alpha _{i}^{2}}}}\psi=\mu\psi~,
\end{equation}
\begin{equation}
{{\sigma }^{(i)}}=\frac{{{\alpha }_{i}}}{\sqrt{1-\sum\limits_{i=1}^{N}{\alpha _{i}^{2}}}}\psi +{{\chi }^{(i)}}=\mu\alpha_i\psi+{{\chi }^{(i)}}~,
\end{equation}
where $\psi$ and  ${{\left\{ {{\chi }^{(i)}} \right\}}_{i=1,2...N}}$ are new scalars, and $\mu=\frac{1}{\sqrt{1-\sum\limits_{i=1}^{N}{\alpha _{i}^{2}}}}$. The Lagrangian becomes
\begin{equation}
\mathcal{L}=-\sum\limits_{i=1}^{N}{\frac{\sqrt{-g}}{2}{{\partial }_{\mu }}{{\chi }^{(i)}}{{\partial }^{\mu }}{{\chi }^{(i)}}}-\frac{\sqrt{-g}}{2}{{\partial }^{\mu }}\psi {{\partial }_{\mu }}\psi -\sqrt{-g}{{V}_{sr}}\left(\mu\psi \right)~.
\end{equation}
The interaction terms of $\chi$ and $\psi$ have been canceled.

The two point correlation function of the inflaton fluctuation can be calculated as
\begin{equation}
\left\langle {{\delta\varphi }_{{{\mathbf{k}}_{\mathbf{1}}}}}{{\delta\varphi }_{{{\mathbf{k}}_{\mathbf{2}}}}} \right\rangle =\frac{{{H}^{2}}}{2k_{1}^{3}}{{\left( 2\pi  \right)}^{3}}{{\delta }^{(3)}}({{\mathbf{k}}_{\mathbf{1}}}+{{\mathbf{k}}_{\mathbf{2}}})\mu=\frac{{{H}^{2}}}{2k_{1}^{3}}{{\left( 2\pi  \right)}^{3}}{{\delta }^{(3)}}({{\mathbf{k}}_{\mathbf{1}}}+{{\mathbf{k}}_{\mathbf{2}}})\frac{1}{1-\sum\limits_{i=1}^{N}{\alpha _{i}^{2}}}~.
\end{equation}
Again, the correction in $\delta\varphi$ is canceled by the correction in $\dot\varphi$ and thus the power spectrum for the curvature perturbation is unchanged if expressed in terms of $H$ and $\epsilon$.

For the bispectrum, at the leading order, it is similar to the $N=1$ case. Namely, the non-Gaussianities are also at $\mathcal{O}{(\epsilon)}$ if no nonlinear interaction is added.
Similar to our previous calculation, the new Lagrangian $\sum\limits_{i=1}^{N}{{{\lambda }_{i}}\varphi A_{\mu }^{(i)}{{\partial }^{\mu }}\varphi }$ still gives trivial power spectrum and bispectrum, up to slow roll suppressed corrections.

Now add interaction
\begin{equation}
\sqrt{-g}\sum\limits_{i=1}^{N}{{{\kappa }_{i}}{{\left( {{A}^{(i)}} \right)}^{4}}}~.
\end{equation}
Decompose similarly to the single vector case, and drop the higher order mixing, we have effectively
\begin{equation}
\sqrt{-g}\sum\limits_{i=1}^{N}{\frac{{{\kappa }_{i}}}{M_{i}^{4}}{{\left( \partial {{\sigma }^{(i)}} \right)}^{4}}}~.
\end{equation}
After rotation, and drop the $\chi$ fields again, we have
\begin{equation}
\sqrt{-g}{{\left( \partial \psi  \right)}^{4}}\left( \sum\limits_{i=1}^{N}{\frac{{{\mu }^{4}}\alpha _{i}^{4}{{\kappa }_{i}}}{M_{i}^{4}}} \right)~.
\end{equation}
Similarly, we write the total effective Lagrangian
\begin{equation}
\mathcal{L}=\sqrt{-g}P(X,\psi )~,
\end{equation}
where
\begin{equation}
X=-\frac{1}{2}{{\left( \partial \psi  \right)}^{2}}~,
\end{equation}
and
\begin{equation}
P=X+4\left( \sum\limits_{i=1}^{N}{\frac{{{\mu }^{4}}\alpha _{i}^{4}{{\kappa }_{i}}}{M_{i}^{4}}} \right){{X}^{2}}-{{V}_{sr}}(\mu \psi )~.
\end{equation}
Now the new $D$ parameter
\begin{equation}
{{D}_{\textrm{tower}}}=4\left( \sum\limits_{i=1}^{N}{\frac{{{\mu }^{4}}\alpha _{i}^{4}{{\kappa }_{i}}}{M_{i}^{4}}} \right)=\frac{4\sum\limits_{i=1}^{N}{\frac{m_{i}^{4}{{\kappa }_{i}}}{M_{i}^{8}}}}{{{\left( 1-\sum\limits_{i=1}^{N}{\frac{m_{i}^{2}}{M_{i}^{2}}} \right)}^{2}}}~.
\end{equation}
Now we can conclude that the mass tower still only changes the parameter $D$ which can be factored out. Now we only consider the case where all $\kappa_i>0$. At the leading order, with choosing suitable tower we can let $\sum\limits_{i=1}^{N}{\frac{m_{i}^{2}}{M_{i}^{2}}}\sim 1$ and cause a dramatically large $DX$, which will drive the leading non-Gaussianities $\left|f_{NL}\right|$ to 0.57 for positive $D_\mathrm{tower}$. For negative $D_\mathrm{tower}$, large non-Gaussianity can be achieved similarly when $D_\mathrm{tower}X$ is near $-1/6$.

\section{Conclusion and discussion}
In this work, we study the coupling between massive vectors and the inflaton field. The model on the one hand can be considered as an extension of quasi-single field inflation, and on the other hand differs in the sense that the actual degree of freedom which couples to the inflaton sector is massless, as long as we restrict our attention to the tree level. Thus equilateral non-Gaussianity, instead of quasi-local family of non-Gaussianities is generated.

We calculate the power spectrum and non-Gaussianity of this model and find that the leading order non-Gaussianity is in general of order one, while having the possibility to become large. Then we consider the multi-vector case. We found that the power spectrum and non-Gaussianity take a similar form to the single vector case.

It is interesting to consider the model beyond the leading order in perturbation theory. Then the divergence free part of vector start to couple to the scalar in loops and may drive the model towards traditional quasi-single field type. It is also interesting to study massive fields with other spins, especially the spin-$1/2$ case. We shall leave them to future work.

\section*{Acknowledgments}
YW was supported by startup funds at the Hong Kong University of Science and Technology, a Starting Grant of the European Research Council (ERC STG grant 279617), and the Stephen Hawking Advanced Fellowship. SZ was supported by the the Hong Kong PhD Fellowship Scheme (HKPFS) issued by the Research Grants Council (RGC) of Hong Kong. JL was supported by the Outstanding Student International Exchange Funding Scheme and Yan Ji-Ci Class in the University of Science and Technology of China. We would also like to thank the Institute for Advanced Study, Hong Kong University of Science and Technology, where this work is finished.

\appendix

\section{The in-in calculation of the power spectrum}
\label{App:thirdorderaction}
\setcounter{equation}{0}

In this part we use the in-in formalism~\cite{Weinberg:2005vy} to compute the two point correlation function of $\varphi$ up to  $\mathcal{O}(\alpha^2)$. For a recent review of the in-in formalism, see~\cite{Chen:2010xka, Wang:2013zva}.
\subsection{Preparation}
We consider the Lagrangian
\begin{equation}
\mathcal{L}={{\mathcal{L}}_{\varphi }}+{{\mathcal{L}}_{\sigma }}+{{\mathcal{L}}_{i}}~,
\end{equation}
where
\begin{equation}
{{\mathcal{L}}_{\varphi }}=-\frac{1}{2}\sqrt{-g}{{\partial }^{\mu }}\delta\varphi {{\partial }_{\mu }}\delta\varphi =\frac{1}{2}{{a}^{3}}\delta\dot{\varphi }\delta\dot{\varphi }-\frac{1}{2}a{{\partial }_{i}}\delta\varphi {{\partial }_{i}}\delta\varphi~,
\end{equation}
\begin{equation}
{{\mathcal{L}}_{\varphi }}=-\frac{1}{2}\sqrt{-g}{{\partial }^{\mu }}\delta\sigma {{\partial }_{\mu }}\delta\sigma =\frac{1}{2}{{a}^{3}}\delta\dot{\sigma }\delta\dot{\sigma }-\frac{1}{2}a{{\partial }_{i}}\delta\sigma {{\partial }_{i}}\delta\sigma~,
\end{equation}
\begin{equation}
{{\mathcal{L}}_{i}}=\sqrt{-g}\frac{m}{M}{{\partial }^{\mu }}\delta\varphi {{\partial }_{\mu }}\delta\sigma =-{{a}^{3}}\frac{m}{M}\delta\dot{\sigma }\delta\dot{\varphi }+a\frac{m}{M}{{\partial }_{i}}\delta\sigma {{\partial }_{i}}\delta\varphi~.
\end{equation}
We define the conjugate momentum as ${{\delta\pi }_{\varphi }}=\frac{\partial \mathcal{L}}{\partial \delta\dot{\varphi }}$ and ${{\delta\pi }_{\sigma }}=\frac{\partial \mathcal{L}}{\partial \delta\dot{\sigma }}$. Then we work out the Hamitonian in terms of $\delta\varphi$, $\delta\sigma$ and $\delta\pi_\varphi$, $\delta\pi_\sigma$ and separate them in to the kinetic part and interaction part.

Now we use the picture transformation to derive the time derivative in the interaction picture (we use the subscript $I$ to denote the configurations in the interaction picture)
\begin{align}
{{\delta\dot{\varphi }}_{I}}=\frac{\partial {{\mathcal{H}}_{0}}}{\partial {{\delta\pi }_{\varphi }}}~,\\
{{\delta\dot{\sigma }}_{I}}=\frac{\partial {{\mathcal{H}}_{0}}}{\partial {{\delta\pi }_{\sigma }}}~.
\end{align}
So we have
\begin{equation}
{{\delta\pi }_{\varphi }}={{a}^{3}}{{\delta\dot{\varphi }}_{I}}~,
\end{equation}
\begin{equation}
{{\delta\pi }_{\sigma }}={{a}^{3}}{{\delta\dot{\sigma }}_{I}}~.
\end{equation}

Finally, we remove the $I$ subscript of fields and plug the inverse mapping into the interaction $\mathcal{H}_i$, we get the interaction Hamiltonian in the interaction picture $\mathcal{H}_I$
\begin{equation}
{{\mathcal{H}}_{I}}={{\mathcal{H}}_{I_1}}+{{\mathcal{H}}_{I_2}}~,
\end{equation}
where
\begin{equation}
\mathcal{H}_{I_1}=\frac{{{a}^{3}}{{\alpha}^{2}}}{2}\delta\dot{\varphi }\delta\dot{\varphi }~,
\end{equation}
\begin{equation}
\mathcal{H}_{I_2}={{a}^{3}}\alpha\delta\dot{\sigma }\delta\dot{\varphi }-a\alpha{{\partial }_{i}}\delta\sigma {{\partial }_{i}}\delta\varphi~.
\end{equation}
The mode functions can be written as
\begin{equation}
{\delta{\varphi }_{\mathbf{k}}}=\delta\varphi _{\mathbf{k}}^{+}+\delta\varphi _{\mathbf{k}}^{-}={{u}_{\mathbf{k}}}(t){{a}_{\mathbf{k}}}+u_{-\mathbf{k}}^{*}(t)a_{-\mathbf{k}}^{\dagger}~,
\end{equation}
\begin{equation}
{\delta{\sigma }_{\mathbf{k}}}=\delta\sigma _{\mathbf{k}}^{+}+\delta\sigma _{\mathbf{k}}^{-}={{v}_{\mathbf{k}}}(t){{b}_{\mathbf{k}}}+v_{-\mathbf{k}}^{*}(t)b_{-\mathbf{k}}^{\dagger}~,
\end{equation}
and the commutation relation
\begin{equation}
[{{a}_{\mathbf{p}}},a_{\mathbf{q}}^{\dagger}]={{(2\pi )}^{3}}{{\delta }^{(3)}}(\mathbf{p}-\mathbf{q})~,
\end{equation}
\begin{equation}
[{{b}_{\mathbf{p}}},b_{\mathbf{q}}^{\dagger}]={{(2\pi )}^{3}}{{\delta }^{(3)}}(\mathbf{p}-\mathbf{q})~.
\end{equation}
The solutions to mode functions are
\begin{equation}
{{u}_{\mathbf{k}}}(\tau )={{v}_{\mathbf{k}}}(\tau )=\frac{H}{\sqrt{2{{k}^{3}}}}(1+ik\tau ){{e}^{-ik\tau }}~,
\end{equation}
where $\tau=-\frac{1}{aH}$. It is convenient to define $x=-k\tau$. Now the mode functions are given by
\begin{equation}
{{u}_{\mathbf{k}}}(x)={{v}_{\mathbf{k}}}(x)=\frac{H}{\sqrt{2{{k}^{3}}}}(1-ix){{e}^{ix}}~.
\end{equation}

Using the in-in formulism, the expectation value for $\delta\varphi^2$ is
\begin{align}
\left\langle {\delta{\varphi }^{2}} \right\rangle &= \frac{H^2}{2k_1^3}(2\pi)^3\delta^{(3)}(\mathbf{k}_\mathbf{1}
+\mathbf{k}_\mathbf{2})+2\operatorname{Im}[\int_{t_0}^{t}{d{t}_1\left\langle 0 \right|\delta\varphi^2\mathcal{H}_{I_1}(t_1)\left| 0 \right\rangle}]\nonumber\\
&+\int_{{{t}_{0}}}^{t}{d{{{\tilde{t}}}_{1}}}\int_{{{t}_{0}}}^{t}{d{{t}_{1}}}\left\langle  0 \right|{{\mathcal{H}}_{I_2}}({{\tilde{t}}_{1}}){{\delta\varphi }^{2}}{{\mathcal{H}}_{I_2}}({{t}_{1}})\left| 0 \right\rangle -2\operatorname{Re}[\int_{{{t}_{0}}}^{t}{d{{t}_{1}}}\int_{{{t}_{0}}}^{{{t}_{1}}}{d{{t}_{2}}}\left\langle  0 \right|{{\delta\varphi }^{2}}{{\mathcal{H}}_{I_2}}({{t}_{1}}){{\mathcal{H}}_{I_2}}({{t}_{2}})\left| 0 \right\rangle ]~,
\end{align}
where ${{\delta\varphi }^{2}}$ is understood as ${{\delta\varphi }_{{{\mathbf{k}}_{\mathbf{1}}}}}(t){{\delta\varphi }_{{{\mathbf{k}}_{\mathbf{2}}}}}(t)$. We can write
\begin{equation}
\left\langle {{\delta\varphi }^{2}} \right\rangle =\frac{H^2}{2k_1^3}(2\pi)^3\delta^{(3)}(\mathbf{k}_\mathbf{1}
+\mathbf{k}_\mathbf{2})+V_0+V_1+V_2~,
\end{equation}
where
\begin{equation}
V_0=2\operatorname{Im}[\int_{t_0}^{t}{d{t}_1\left\langle 0 \right|\delta\varphi^2\mathcal{H}_{I_1}(t_1)\left| 0 \right\rangle}]~,
\end{equation}
\begin{equation}
V_1=\int_{{{t}_{0}}}^{t}{d{{{\tilde{t}}}_{1}}}\int_{{{t}_{0}}}^{t}{d{{t}_{1}}}\left\langle  0 \right|{{\mathcal{H}}_{I_2}}({{\tilde{t}}_{1}}){{\delta\varphi }^{2}}{{\mathcal{H}}_{I_2}}({{t}_{1}})\left| 0 \right\rangle~,
\end{equation}
\begin{equation}
V_2=-2\operatorname{Re}[\int_{{{t}_{0}}}^{t}{d{{t}_{1}}}\int_{{{t}_{0}}}^{{{t}_{1}}}{d{{t}_{2}}}\left\langle  0 \right|{{\delta\varphi }^{2}}{{\mathcal{H}}_{I_2}}({{t}_{1}}){{\mathcal{H}}_{I_2}}({{t}_{2}})\left| 0 \right\rangle ]~.
\end{equation}
\subsection{Calculation of $V_0$}
In this section we calculate
\begin{equation}
V_0=2\operatorname{Im}[\int_{t_0}^{t}{d{t}_1\left\langle 0 \right|\delta\varphi^2\mathcal{H}_{I_1}(t_1)\left| 0 \right\rangle}]~.
\end{equation}

Using the momentum version of Hamiltonian, we have
\begin{equation}
{{V}_{0}}=\frac{{{m}^{2}}}{{{M}^{2}}}\operatorname{Im}\int_{{{t}_{0}}}^{t}{d{{t}_{1}}}\left\langle  0 \right|{{\delta\varphi }_{{{\mathbf{k}}_{\mathbf{1}}}}}(t){{\delta\varphi }_{{{\mathbf{k}}_{\mathbf{2}}}}}(t){{a}^{3}}({{t}_{1}}){{\delta\dot{\varphi }}_{\mathbf{p}}}({{t}_{1}}){{\delta\dot{\varphi }}_{-\mathbf{p}}}({{t}_{1}})\left| 0 \right\rangle~.
\end{equation}

We use the Wick contraction to deal with the integration. The integration can be represented as two kinds of contractions because of the rearrangement of ${{\delta\varphi }_{{{\mathbf{k}}_{\mathbf{1}}}}}(t)$ and ${{\delta\varphi }_{{{\mathbf{k}}_{\mathbf{2}}}}}(t)$. If the integration is divergent, we can use the $i\epsilon$ prescription and the cutoff (which means we can choose the lower limit of integrations to be $e^{-N_e}$ instead of $0$, where $N_e$ is the efolding number in the inflationary process) to get a finite value. One can derive that
\begin{equation}
{{V}_{0}}=\frac{{{m}^{2}}{{H}^{2}}}{{2{M}^{2}}k_{1}^{3}}{{(2\pi )}^{3}}{{\delta }^{(3)}}({{\mathbf{k}}_{\mathbf{1}}}+{{\mathbf{k}}_{\mathbf{2}}})\operatorname{Im}\int_{0}^{+\infty }{d{{x}_{1}}{{e}^{-2i{{x}_{1}}}}}~.
\end{equation}
After $i\epsilon$ process we have
\begin{equation}
\int_{0}^{+\infty }{d{{x}_{1}}{{e}^{-2i{{x}_{1}}}}}=-\frac{i}{2}~.
\end{equation}
So the result is
\begin{equation}
{{V}_{0}}=-\frac{{{m}^{2}}{{H}^{2}}}{4{{M}^{2}}k_{1}^{3}}{{(2\pi )}^{3}}{{\delta }^{(3)}}({{\mathbf{k}}_{\mathbf{1}}}+{{\mathbf{k}}_{\mathbf{2}}})~.
\end{equation}
\subsection{Calculation of $V_1$}
In this section we calculate
\begin{equation}
V_1=\int_{{{t}_{0}}}^{t}{d{{{\tilde{t}}}_{1}}}\int_{{{t}_{0}}}^{t}{d{{t}_{1}}}\left\langle  0 \right|{{\mathcal{H}}_{I_2}}({{\tilde{t}}_{1}}){{\delta\varphi }^{2}}{{\mathcal{H}}_{I_2}}({{t}_{1}})\left| 0 \right\rangle~.
\end{equation}

Use the momentum version of Hamiltonian, we have
\begin{align}
&{{V}_{1}}=\int_{{{t}_{0}}}^{t}{d{{{\tilde{t}}}_{1}}}\int_{{{t}_{0}}}^{t}{d{{t}_{1}}}\left\langle  0 \right|({{a}^{3}}({{\tilde{t}}_{1}})\frac{m}{M}{{\delta\dot{\sigma }}_{\mathbf{p}}}({{\tilde{t}}_{1}}){{\delta\dot{\varphi }}_{-\mathbf{p}}}({{\tilde{t}}_{1}})-a({{\tilde{t}}_{1}})\frac{m}{M}{{p}^{2}}{{\delta\sigma }_{\mathbf{p}}}({{\tilde{t}}_{1}}){{\delta\varphi }_{-\mathbf{p}}}({{\tilde{t}}_{1}})){{\delta\varphi }_{{{\mathbf{k}}_{\mathbf{1}}}}}(t){{\delta\varphi }_{{{\mathbf{k}}_{\mathbf{2}}}}}(t)\\
&({{a}^{3}}({{t}_{1}})\frac{m}{M}{{\delta\dot{\sigma }}_{\mathbf{q}}}({{t}_{1}}){{\delta\dot{\varphi }}_{-\mathbf{q}}}({{t}_{1}})-a({{t}_{1}})\frac{m}{M}{{q}^{2}}{{\delta\sigma }_{\mathbf{q}}}({{t}_{1}}){{\delta\varphi }_{-\mathbf{q}}}({{t}_{1}}))\left| 0 \right\rangle~.
\end{align}
Split it into four terms, we have
\begin{equation}
V_1=V_{11}+V_{12}+V_{13}+V_{14}~,
\end{equation}
where
\begin{equation}
{{V}_{11}}=\frac{{{m}^{2}}}{{{M}^{2}}}\int_{{{t}_{0}}}^{t}{d{{{\tilde{t}}}_{1}}}\int_{{{t}_{0}}}^{t}{d{{t}_{1}}}\left\langle  0 \right|{{a}^{3}}({{\tilde{t}}_{1}}){{\delta\dot{\sigma }}_{\mathbf{p}}}({{\tilde{t}}_{1}}){{\delta\dot{\varphi }}_{-\mathbf{p}}}({{\tilde{t}}_{1}}){{\delta\varphi }_{{{\mathbf{k}}_{\mathbf{1}}}}}(t){{\delta\varphi }_{{{\mathbf{k}}_{\mathbf{2}}}}}(t){{a}^{3}}({{t}_{1}}){{\delta\dot{\sigma }}_{\mathbf{q}}}({{t}_{1}}){{\delta\dot{\varphi }}_{-\mathbf{q}}}({{t}_{1}})\left| 0 \right\rangle~,
\end{equation}
\begin{equation}
{{V}_{12}}=-\frac{{{m}^{2}}}{{{M}^{2}}}\int_{{{t}_{0}}}^{t}{d{{{\tilde{t}}}_{1}}}\int_{{{t}_{0}}}^{t}{d{{t}_{1}}}\left\langle  0 \right|{{a}^{3}}({{\tilde{t}}_{1}}){{\delta\dot{\sigma }}_{\mathbf{p}}}({{\tilde{t}}_{1}}){{\delta\dot{\varphi }}_{-\mathbf{p}}}({{\tilde{t}}_{1}}){{\delta\varphi }_{{{\mathbf{k}}_{\mathbf{1}}}}}(t){{\delta\varphi }_{{{\mathbf{k}}_{\mathbf{2}}}}}(t)a({{t}_{1}}){{q}^{2}}{{\delta\sigma }_{\mathbf{q}}}({{t}_{1}}){{\delta\varphi }_{-\mathbf{q}}}({{t}_{1}})\left| 0 \right\rangle~,
\end{equation}
\begin{equation}
{{V}_{13}}=-\frac{{{m}^{2}}}{{{M}^{2}}}\int_{{{t}_{0}}}^{t}{d{{{\tilde{t}}}_{1}}}\int_{{{t}_{0}}}^{t}{d{{t}_{1}}}\left\langle  0 \right|a({{\tilde{t}}_{1}}){{p}^{2}}{{\delta\sigma }_{\mathbf{p}}}({{\tilde{t}}_{1}}){{\delta\varphi }_{-\mathbf{p}}}({{\tilde{t}}_{1}}){{\delta\varphi }_{{{\mathbf{k}}_{\mathbf{1}}}}}(t){{\delta\varphi }_{{{\mathbf{k}}_{\mathbf{2}}}}}(t){{a}^{3}}({{t}_{1}}){{\delta\dot{\sigma }}_{\mathbf{q}}}({{t}_{1}}){{\delta\dot{\varphi }}_{-\mathbf{q}}}({{t}_{1}})\left| 0 \right\rangle~,
\end{equation}
\begin{equation}
{{V}_{14}}=\frac{{{m}^{2}}}{{{M}^{2}}}\int_{{{t}_{0}}}^{t}{d{{{\tilde{t}}}_{1}}}\int_{{{t}_{0}}}^{t}{d{{t}_{1}}}\left\langle  0 \right|a({{\tilde{t}}_{1}}){{p}^{2}}{{\delta\sigma }_{\mathbf{p}}}({{\tilde{t}}_{1}}){{\delta\varphi }_{-\mathbf{p}}}({{\tilde{t}}_{1}}){{\delta\varphi }_{{{\mathbf{k}}_{\mathbf{1}}}}}(t){{\delta\varphi }_{{{\mathbf{k}}_{\mathbf{2}}}}}(t)a({{t}_{1}}){{q}^{2}}{{\delta\sigma }_{\mathbf{q}}}({{t}_{1}}){{\delta\varphi }_{-\mathbf{q}}}({{t}_{1}})\left| 0 \right\rangle~.
\end{equation}

For $V_{11}$, we have
\begin{equation}
{{V}_{11}}=\frac{{{m}^{2}}{{H}^{2}}}{4{{M}^{2}}k_{1}^{3}}{{(2\pi )}^{3}}{{\delta }^{(3)}}({{\mathbf{k}}_{\mathbf{1}}}+{{\mathbf{k}}_{\mathbf{2}}}){{\left| \int_{0}^{+\infty }{d{{x}_{1}}{{e}^{2i{{x}_{1}}}}} \right|}^{2}}~.
\end{equation}
With the $i\epsilon$ prescription, we have
\begin{equation}
\int_{0}^{+\infty }{d{{x}_{1}}{{e}^{2i{{x}_{1}}}}}=\frac{i}{2}~.
\end{equation}
So
\begin{equation}
{{V}_{11}}=\frac{{{m}^{2}}{{H}^{2}}}{16{{M}^{2}}k_{1}^{3}}{{(2\pi )}^{3}}{{\delta }^{(3)}}({{\mathbf{k}}_{\mathbf{1}}}+{{\mathbf{k}}_{\mathbf{2}}})~.
\end{equation}

For $V_{12}$, similarly we have
\begin{equation}
{{V}_{12}}=-\frac{{{m}^{2}}{{H}^{2}}}{4{{M}^{2}}k_{1}^{3}}{{(2\pi )}^{3}}{{\delta }^{(3)}}({{\mathbf{k}}_{\mathbf{1}}}+{{\mathbf{k}}_{\mathbf{2}}})\int_{0}^{+\infty }{d{{{\tilde{x}}}_{1}}}{{e}^{2i{{{\tilde{x}}}_{1}}}}\int_{0}^{+\infty }{d{{x}_{1}}}\frac{{{(1+i{{x}_{1}})}^{2}}}{x_{1}^{2}}{{e}^{-2i{{x}_{1}}}}~.
\end{equation}
Similarly, this integration can be evaluated using the $i\epsilon$ prescription,
\begin{equation}
{{V}_{12}}=-\frac{{{m}^{2}}{{H}^{2}}}{4{{M}^{2}}k_{1}^{3}}{{(2\pi )}^{3}}{{\delta }^{(3)}}({{\mathbf{k}}_{\mathbf{1}}}+{{\mathbf{k}}_{\mathbf{2}}})(\frac{i}{2}{{e}^{N_e}}+\frac{3}{4})~.
\end{equation}

It is very easy to show that
\begin{equation}
{{V}_{13}}=V_{12}^{*}~.
\end{equation}
So we have
\begin{equation}
{{V}_{12}}+{{V}_{13}}=2\operatorname{Re}{{V}_{12}}=-\frac{3{{m}^{2}}{{H}^{2}}}{8{{M}^{2}}k_{1}^{3}}{{(2\pi )}^{3}}{{\delta }^{(3)}}({{\mathbf{k}}_{\mathbf{1}}}+{{\mathbf{k}}_{\mathbf{2}}})~.
\end{equation}

For $V_{14}$, similarly
\begin{equation}
{{V}_{14}}=\frac{{{m}^{2}}{{H}^{2}}}{4{{M}^{2}}k_{1}^{3}}{{(2\pi )}^{3}}{{\delta }^{(3)}}({{\mathbf{k}}_{\mathbf{1}}}+{{\mathbf{k}}_{\mathbf{2}}}){{\left| \int_{0}^{+\infty }{d{{x}_{1}}}\frac{{{(1+i{{x}_{1}})}^{2}}}{x_{1}^{2}}{{e}^{-2i{{x}_{1}}}} \right|}^{2}}~.
\end{equation}
Using the $i\epsilon$ prescription to evaluate the integrals, we have
\begin{equation}
{{V}_{14}}=\frac{{{m}^{2}}{{H}^{2}}}{{{M}^{2}}k_{1}^{3}}{{(2\pi )}^{3}}{{\delta }^{(3)}}({{\mathbf{k}}_{\mathbf{1}}}+{{\mathbf{k}}_{\mathbf{2}}})(\frac{9}{16}+\frac{1}{4}{{e}^{2N_e}})~.
\end{equation}

Summing up those contributions, we obtain $V_1$ as
\begin{align}
{{V}_{1}}&={{V}_{11}}+{{V}_{12}}+{{V}_{13}}+{{V}_{14}}\nonumber\\
&=\frac{{{m}^{2}}{{H}^{2}}}{{{M}^{2}}k_{1}^{3}}{{(2\pi )}^{3}}{{\delta }^{(3)}}({{\mathbf{k}}_{\mathbf{1}}}+{{\mathbf{k}}_{\mathbf{2}}})(\frac{1}{4}+\frac{1}{4}{{e}^{2N_e}})~.
\end{align}
\subsection{Calculation of $V_2$}
In this section we calculate
\begin{equation}
V_2=-2\operatorname{Re}[\int_{{{t}_{0}}}^{t}{d{{t}_{1}}}\int_{{{t}_{0}}}^{{{t}_{1}}}{d{{t}_{2}}}\left\langle  0 \right|{{\delta\varphi }^{2}}{{\mathcal{H}}_{I_2}}({{t}_{1}}){{\mathcal{H}}_{I_2}}({{t}_{2}})\left| 0 \right\rangle ]~.
\end{equation}

Similarly, plug in the explicit form of Hamiltonian, we have
\begin{align}
&{{V}_{2}}=-2\operatorname{Re}\int_{{{t}_{0}}}^{t}{d{{t}_{1}}}\int_{{{t}_{0}}}^{{{t}_{1}}}{d{{t}_{2}}}\left\langle  0 \right|{{\delta\varphi }_{{{\mathbf{k}}_{\mathbf{1}}}}}(t){{\delta\varphi }_{{{\mathbf{k}}_{\mathbf{2}}}}}(t)({{a}^{3}}({{t}_{1}})\frac{m}{M}{{\delta\dot{\sigma }}_{\mathbf{p}}}({{t}_{1}}){{\delta\dot{\varphi }}_{-\mathbf{p}}}({{t}_{1}})-a({{t}_{1}})\frac{m}{M}{{p}^{2}}{{\delta\sigma }_{\mathbf{p}}}({{t}_{1}}){{\delta\varphi }_{-\mathbf{p}}}({{t}_{1}}))\nonumber\\
&({{a}^{3}}({{t}_{2}})\frac{m}{M}{{\delta\dot{\sigma }}_{\mathbf{q}}}({{t}_{2}}){{\delta\dot{\varphi }}_{-\mathbf{q}}}({{t}_{2}})-a({{t}_{2}})\frac{m}{M}{{q}^{2}}{{\delta\sigma }_{\mathbf{q}}}({{t}_{2}}){{\delta\varphi }_{-\mathbf{q}}}({{t}_{2}}))\left| 0 \right\rangle~.
\end{align}
Split it as the last section, we have
\begin{equation}
{{V}_{2}}={{V}_{21}}+{{V}_{22}}+{{V}_{23}}+{{V}_{24}}~,
\end{equation}
where
\begin{equation}
{{V}_{21}}=-2\frac{{{m}^{2}}}{{{M}^{2}}}\operatorname{Re}\int_{{{t}_{0}}}^{t}{d{{t}_{1}}}\int_{{{t}_{0}}}^{{{t}_{1}}}{d{{t}_{2}}}\left\langle  0 \right|{{\delta\varphi }_{{{\mathbf{k}}_{\mathbf{1}}}}}(t){{\delta\varphi }_{{{\mathbf{k}}_{\mathbf{2}}}}}(t){{a}^{3}}({{t}_{1}}){{\delta\dot{\sigma }}_{\mathbf{p}}}({{t}_{1}}){{\delta\dot{\varphi }}_{-\mathbf{p}}}({{t}_{1}}){{a}^{3}}({{t}_{2}}){{\delta\dot{\sigma }}_{\mathbf{q}}}({{t}_{2}}){{\delta\dot{\varphi }}_{-\mathbf{q}}}({{t}_{2}})\left| 0 \right\rangle~,
\end{equation}
\begin{equation}
{{V}_{22}}=2\frac{{{m}^{2}}}{{{M}^{2}}}\operatorname{Re}\int_{{{t}_{0}}}^{t}{d{{t}_{1}}}\int_{{{t}_{0}}}^{{{t}_{1}}}{d{{t}_{2}}}\left\langle  0 \right|{{\delta\varphi }_{{{\mathbf{k}}_{\mathbf{1}}}}}(t){{\delta\varphi }_{{{\mathbf{k}}_{\mathbf{2}}}}}(t){{a}^{3}}({{t}_{1}}){{\delta\dot{\sigma }}_{\mathbf{p}}}({{t}_{1}}){{\delta\dot{\varphi }}_{-\mathbf{p}}}({{t}_{1}})a({{t}_{2}}){{q}^{2}}{{\delta\sigma }_{\mathbf{q}}}({{t}_{2}}){{\delta\varphi }_{-\mathbf{q}}}({{t}_{2}})\left| 0 \right\rangle~,
\end{equation}
\begin{equation}
{{V}_{23}}=2\frac{{{m}^{2}}}{{{M}^{2}}}\operatorname{Re}\int_{{{t}_{0}}}^{t}{d{{t}_{1}}}\int_{{{t}_{0}}}^{{{t}_{1}}}{d{{t}_{2}}}\left\langle  0 \right|{{\delta\varphi }_{{{\mathbf{k}}_{\mathbf{1}}}}}(t){{\delta\varphi }_{{{\mathbf{k}}_{\mathbf{2}}}}}(t)a({{t}_{1}}){{p}^{2}}{{\delta\sigma }_{\mathbf{p}}}({{t}_{1}}){{\delta\varphi }_{-\mathbf{p}}}({{t}_{1}}){{a}^{3}}({{t}_{2}}){{\delta\dot{\sigma }}_{\mathbf{q}}}({{t}_{2}}){{\delta\dot{\varphi }}_{-\mathbf{q}}}({{t}_{2}})\left| 0 \right\rangle~,
\end{equation}
\begin{equation}
{{V}_{24}}=-2\frac{{{m}^{2}}}{{{M}^{2}}}\operatorname{Re}\int_{{{t}_{0}}}^{t}{d{{t}_{1}}}\int_{{{t}_{0}}}^{{{t}_{1}}}{d{{t}_{2}}}\left\langle  0 \right|{{\delta\varphi }_{{{\mathbf{k}}_{\mathbf{1}}}}}(t){{\delta\varphi }_{{{\mathbf{k}}_{\mathbf{2}}}}}(t)a({{t}_{1}}){{p}^{2}}{{\delta\sigma }_{\mathbf{p}}}({{t}_{1}}){{\delta\varphi }_{-\mathbf{p}}}({{t}_{1}})a({{t}_{2}}){{q}^{2}}{{\delta\sigma }_{\mathbf{q}}}({{t}_{2}}){{\delta\varphi }_{-\mathbf{q}}}({{t}_{2}})\left| 0 \right\rangle~.
\end{equation}

For $V_{21}$, using the same technologies in the last section, we have
\begin{equation}
{{V}_{21}}=\frac{{{m}^{2}}{{H}^{2}}}{8{{M}^{2}}k_{1}^{3}}{{(2\pi )}^{3}}{{\delta }^{(3)}}({{\mathbf{k}}_{\mathbf{1}}}+{{\mathbf{k}}_{\mathbf{2}}})~.
\end{equation}

Similarly, for $V_{22}$
\begin{equation}
{{V}_{22}}=\frac{{{m}^{2}}{{H}^{2}}}{2{{M}^{2}}k_{1}^{3}}{{(2\pi )}^{3}}{{\delta }^{(3)}}({{\mathbf{k}}_{\mathbf{1}}}+{{\mathbf{k}}_{\mathbf{2}}})\operatorname{Re}\int_{0}^{+\infty }{d{{x}_{1}}}\int_{{{x}_{1}}}^{+\infty }{d{{x}_{2}}\frac{{{(1+i{{x}_{2}})}^{2}}}{x_{2}^{2}}}{{e}^{-2i{{x}_{2}}}}~.
\end{equation}
It is not hard to evaluate the integration
\begin{equation}
{{V}_{22}}=\frac{{{m}^{2}}{{H}^{2}}}{{{M}^{2}}k_{1}^{3}}{{(2\pi )}^{3}}{{\delta }^{(3)}}({{\mathbf{k}}_{\mathbf{1}}}+{{\mathbf{k}}_{\mathbf{2}}})(-\frac{1}{2}\log 2-\frac{1}{2}{{\gamma }_{E}}+\frac{1}{2}N_e+\frac{1}{8})~,
\end{equation}
where ${\gamma }_{E}$ is the Euler $\gamma$ constant, the $N_e$ is the e-folding number.

$V_{23}$ is very similar to $V_{22}$. One can derive that
\begin{equation}
{{V}_{23}}=\frac{{{m}^{2}}{{H}^{2}}}{2{{M}^{2}}k_{1}^{3}}{{(2\pi )}^{3}}{{\delta }^{(3)}}({{\mathbf{k}}_{\mathbf{1}}}+{{\mathbf{k}}_{\mathbf{2}}})\operatorname{Re}\int_{0}^{+\infty }{d{{x}_{1}}}\int_{{{x}_{1}}}^{+\infty }{d{{x}_{2}}(1+\frac{1}{x_{1}^{2}})}{{e}^{-2i{{x}_{2}}}}~.
\end{equation}
Evaluating the integration as
\begin{equation}
{{V}_{23}}=\frac{{{m}^{2}}{{H}^{2}}}{{{M}^{2}}k_{1}^{3}}{{(2\pi )}^{3}}{{\delta }^{(3)}}({{\mathbf{k}}_{\mathbf{1}}}+{{\mathbf{k}}_{\mathbf{2}}})(\frac{1}{2}\log 2+\frac{1}{2}{{\gamma }_{E}}-\frac{1}{2}N_e-\frac{5}{8})~,
\end{equation}
we find that
\begin{equation}
{{V}_{22}}+{{V}_{23}}=-\frac{{{m}^{2}}{{H}^{2}}}{2{{M}^{2}}k_{1}^{3}}{{(2\pi )}^{3}}{{\delta }^{(3)}}({{\mathbf{k}}_{\mathbf{1}}}+{{\mathbf{k}}_{\mathbf{2}}})~.
\end{equation}

For $V_{24}$, we have
\begin{equation}
{{V}_{24}}=-\frac{{{m}^{2}}{{H}^{2}}}{2{{M}^{2}}k_{1}^{3}}{{(2\pi )}^{3}}{{\delta }^{(3)}}({{\mathbf{k}}_{\mathbf{1}}}+{{\mathbf{k}}_{\mathbf{2}}})\operatorname{Re}\int_{0}^{+\infty }{d{{x}_{1}}}\int_{{{x}_{1}}}^{+\infty }{d{{x}_{2}}\frac{{{(1+{{x}_{1}})}^{2}}}{x_{1}^{2}}\frac{{{(1+i{{x}_{2}})}^{2}}}{x_{2}^{2}}}{{e}^{-2i{{x}_{2}}}}~.
\end{equation}
Evaluating the integration as
\begin{equation}
{{V}_{24}}=\frac{{{m}^{2}}{{H}^{2}}}{{{M}^{2}}k_{1}^{3}}{{(2\pi )}^{3}}{{\delta }^{(3)}}({{\mathbf{k}}_{\mathbf{1}}}+{{\mathbf{k}}_{\mathbf{2}}})(\frac{7}{8}-\frac{1}{4}{{e}^{2N_e}})~.
\end{equation}

As a summary, we can derive that
\begin{align}
{{V}_{2}}&={{V}_{21}}+{{V}_{22}}+{{V}_{23}}+{{V}_{24}}\nonumber\\
&=\frac{{{m}^{2}}{{H}^{2}}}{{{M}^{2}}k_{1}^{3}}{{(2\pi )}^{3}}{{\delta }^{(3)}}({{\mathbf{k}}_{\mathbf{1}}}+{{\mathbf{k}}_{\mathbf{2}}})(\frac{1}{2}-\frac{1}{4}{{e}^{2N_e}})~.
\end{align}
\subsection{The final result}
We can see that the singular terms in $V_1$ and $V_2$ are canceled. Now we can see
\begin{align}
\left\langle {{\varphi }^{2}} \right\rangle &=\frac{H^2}{2k_1^3}(2\pi)^3\delta^{(3)}(\mathbf{k}_\mathbf{1}
+\mathbf{k}_\mathbf{2})+V_0+V_1+V_2\nonumber\\
&=\frac{{{{H}^{2}}}}{2k_{1}^{3}}{{(2\pi )}^{3}}{{\delta}^{(3)}}({{\mathbf{k}}_{\mathbf{1}}}+{{\mathbf{k}}_{\mathbf{2}}})(1+\frac{m^2}{M^2})~,
\end{align}
which agrees with the leading result from field space rotation.


\begin{thebibliography}{999}
\bibitem{Guth:1980zm}
  A.~H.~Guth,
  Phys.\ Rev.\ D {\bf 23}, 347 (1981).

\bibitem{Linde:1981mu}
  A.~D.~Linde,
  Phys.\ Lett.\ B {\bf 108}, 389 (1982).

\bibitem{Albrecht:1982wi}
  A.~Albrecht and P.~J.~Steinhardt,
  Phys.\ Rev.\ Lett.\  {\bf 48}, 1220 (1982).

\bibitem{Chen:2009we}
  X.~Chen and Y.~Wang,
  Phys.\ Rev.\ D {\bf 81}, 063511 (2010)
  [arXiv:0909.0496 [astro-ph.CO]].

\bibitem{Chen:2009zp}
  X.~Chen and Y.~Wang,
  JCAP {\bf 1004}, 027 (2010)
  [arXiv:0911.3380 [hep-th]].

\bibitem{Chen:2012ge}
  X.~Chen and Y.~Wang,
  JCAP {\bf 1209}, 021 (2012)
  [arXiv:1205.0160 [hep-th]].

\bibitem{Pi:2012gf}
  S.~Pi and M.~Sasaki,
  JCAP {\bf 1210}, 051 (2012)
  [arXiv:1205.0161 [hep-th]].

\bibitem{Green:2013rd}
  D.~Green, M.~Lewandowski, L.~Senatore, E.~Silverstein and M.~Zaldarriaga,
  JHEP {\bf 1310}, 171 (2013)
  [arXiv:1301.2630].
  
\bibitem{Ford:1989me}
  L.~H.~Ford,
  Phys.\ Rev.\ D {\bf 40}, 967 (1989).

\bibitem{Maleknejad:2011jw}
  A.~Maleknejad and M.~M.~Sheikh-Jabbari,
  Phys.\ Lett.\ B {\bf 723}, 224 (2013)
  [arXiv:1102.1513 [hep-ph]].
  
\bibitem{Solomon:2013iza}
  A.~R.~Solomon and J.~D.~Barrow,
  Phys.\ Rev.\ D {\bf 89}, no. 2, 024001 (2014)
  [arXiv:1309.4778 [astro-ph.CO]].

\bibitem{Baumann:2011nk}
  D.~Baumann and D.~Green,
  Phys.\ Rev.\ D {\bf 85}, 103520 (2012)
  [arXiv:1109.0292 [hep-th]].

\bibitem{Sasaki:2008uc}
  M.~Sasaki,
  Prog.\ Theor.\ Phys.\  {\bf 120}, 159 (2008)
  [arXiv:0805.0974 [astro-ph]].

\bibitem{Huang:2009vk}
  Q.~G.~Huang,
  JCAP {\bf 0906}, 035 (2009)
  [arXiv:0904.2649 [hep-th]].

\bibitem{Chen:2006nt}
  X.~Chen, M.~x.~Huang, S.~Kachru and G.~Shiu,
  JCAP {\bf 0701}, 002 (2007)
  [hep-th/0605045].

\bibitem{Weinberg:2005vy}
  S.~Weinberg,
  Phys.\ Rev.\ D {\bf 72}, 043514 (2005)
  [hep-th/0506236].

\bibitem{Chen:2010xka}
  X.~Chen,
  Adv.\ Astron.\  {\bf 2010}, 638979 (2010)
  [arXiv:1002.1416 [astro-ph.CO]].

\bibitem{Wang:2013zva}
  Y.~Wang,
  Commun.\ Theor.\ Phys.\  {\bf 62}, 109 (2014)
  [arXiv:1303.1523 [hep-th]].

\end{thebibliography}
\end{document}